\documentclass{article}
\usepackage[T1]{fontenc} 
\usepackage[utf8]{inputenc} 
\usepackage{ismir,amsmath,cite,url}
\usepackage{graphicx}
\usepackage{xcolor}
\usepackage{glossaries}
\usepackage{xspace}
\usepackage{amssymb}
\usepackage{booktabs}
\usepackage[bookmarks=false,hidelinks]{hyperref}
\usepackage{musicography}
\usepackage[linesnumbered,ruled,vlined]{algorithm2e}
\usepackage{microtype}

\usepackage{lineno}

\newcommand{\X}{\mathcal{X}}
\newcommand{\Hid}{\mathcal{H}}
\newcommand{\head}{\textrm{head}}
\newcommand{\dep}{\textrm{dep}}
\newcommand{\level}{l}
\newcommand{\potarcs}{\Lambda}
\newcommand{\nfeat}{\phi}
\newcommand{\len}{\lambda}

\def\ourapp{\textsf{\small MuDeP}\xspace}

\def\<#1>{\langle #1 \rangle}

\def\Music21{\textsf{Music21}\xspace}

\def\ps13{\textsf{\small ps13}\xspace}

\def\Reduction#1#2#3#4{%
\mathrel{\raise1.0ex\hbox{%
\vtop{\ialign{##\crcr%
\raise0.0ex\hbox{$\hfil\scriptstyle{\ #1\ }\hfil$}\crcr%
\noalign{\nointerlineskip}%
\rightarrowfill\crcr%
\noalign{\nointerlineskip}%
\raise0.0ex\hbox{$\hfil\scriptstyle{\ #2\ }\hfil$}\crcr}}}{}^{#3}_{#4}}}
\def\Leduction#1#2#3#4{%
\mathrel{\raise1.0ex\hbox{%
\vtop{\ialign{##\crcr%
\raise0.0ex\hbox{$\hfil\scriptstyle{\ #1\ }\hfil$}\crcr%
\noalign{\nointerlineskip}%
\leftarrowfill\crcr%
\noalign{\nointerlineskip}%
$\hfil\scriptstyle{\ #2\ }\hfil$\crcr}}}{}^{#3}_{#4}}}
\def\hookReduction#1#2#3#4{%
\mathrel{\raise1.2ex\hbox{%
\vtop{\ialign{##\crcr%
\raise0.0ex\hbox{$\hfil\scriptstyle{\ #1\ }\hfil$}\crcr%
\noalign{\nointerlineskip}%
$\lhook\joinrel$\hspace{-0.35em}
\rightarrowfill\crcr%
\noalign{\nointerlineskip}%
$\hfil\scriptstyle{\ #2\ }\hfil$\crcr}}}{}^{#3}_{#4}}}
\def\hoookReduction#1#2#3#4{%
\lhook\joinrel\hspace{-0.50em}
\raise0.85ex\hbox{%
\vtop{\ialign{##\crcr%
\raise0.4ex\hbox{$\hfil\scriptstyle{\ #1\ }\hfil$}\crcr%
\noalign{\nointerlineskip}%
\rightarrowfill\crcr%
\noalign{\nointerlineskip}%
$\hfil\scriptstyle{\ #2\ }\hfil$\crcr}}}{}^{#3}_{#4}}

\def\frew#1#2#3#4#5#6#7#8{
\setbox0=\hbox{$#6 #7 #1 #8$}%
\setbox1=\hbox{$#6 #7 #2 #8$}%
\ifdim \wd0>\wd1 \rlap{\rlap{\hbox to \wd0{#5}}%
                            {\hbox to\wd0{\hfil\lower #3\box1\relax\hfil}}}{\raise #4\box0}%
\else \rlap{\rlap{\hbox to \wd1{#5}}{\hbox to\wd1{\hfil\raise #4\box0\relax\hfil}}}{\lower #3\box1}%
\fi
}

\title{Predicting Music Hierarchies With a \\ Graph-Based Neural Decoder}

\oneauthor
{Francesco Foscarin$^{1}$ \hspace{1cm} Daniel Harasim$^{2}$ \hspace{1cm}  Gerhard Widmer$^{1}$} {$^1$ Institute of Computational Perception, Johannes Kepler University Linz, Austria \\ $^2$ Independent Researcher \\ {\tt francesco.foscarin@jku.at}}

\def\authorname{F. Foscarin, D. Harasim, and G. Widmer}

\begin{document}

\maketitle
\begin{abstract}
This paper describes a data-driven framework to parse musical sequences into dependency trees, which are hierarchical structures used in music cognition research and music analysis.
The parsing involves two steps. First, the input sequence is passed through a transformer encoder to enrich it with contextual information. 
Then, a classifier filters the graph of all possible dependency arcs to produce the dependency tree.
One major benefit of this system is that it can be easily integrated into modern deep-learning pipelines. Moreover, since it does not rely on any particular symbolic grammar, it can consider multiple musical features simultaneously, make use of sequential context information, and produce partial results for noisy inputs.
We test our approach on two datasets of musical trees -- time-span trees of monophonic note sequences and harmonic trees of jazz chord sequences -- and show that our approach outperforms previous methods.\footnote{All our code and data are publicly available at \url{https://github.com/fosfrancesco/musicparser}}

\end{abstract}


\section{Introduction}\label{sec:introduction}

Tree-like representations are a powerful tool in many approaches to music analysis, such as Schenkerian Theory and the Generative Theory of Tonal Music (GTTM). In the Music Information Retrieval (MIR) literature, we find tree models of
melodies \cite{abdallah2015analysing,nakamura2016tree,hamanakatime,finkensiep2021modeling}, chord progressions \cite{rohrmeier2011towards,granroth2014robust,harasim2018generalized, melkonian2019music}, and rhythm \cite{harasim2019harmonic, foscarin2019modeling,foscarin2019parse,foscarin2019diff,rohrmeier2020towards}. 
Parallels between aspects of music and language are often drawn, as these have similar hierarchical properties and their underlying cognitive mechanisms could be closely related~\cite{fitch2014hierarchical}.
However, with a few exceptions, such as 
instrument grouping and metrical information in scores, music is generally encoded sequentially without explicit information about its hierarchical organisation. The task of creating such hierarchies from a sequential representation is called \textit{parsing} and it is an active object of study in the MIR community~\cite{tsuchiya2013probabilistic,harasim2018generalized,foscarin2019parse,hamanakatime}.

Current parsing approaches are based on \textit{generative grammars}, typically context-free-grammars (CFG) or similar related mechanisms, which can be fundamentally seen as a set of expansion rules generating a tree
from the top (the root) to the elements that compose the sequence (the leaves). 
Grammar rules can be enriched with a probability model that permits the ranking of different parses by plausibility. When a grammar is available, parsing can be achieved with grammar-based parsing algorithms, typically variants of the Cocke–Younger–Kasami (CYK) algorithm~\cite{sakai1961syntax}.
While the grammar rules are most often built by hand, by relying on musicologists' knowledge, the probabilities can be learnt from data if sufficient amounts of musical sequences with ground-truth tree annotations are available.
The grammar approach has the strong advantage of leveraging an interpretable and cognitively plausible mechanism. Still, it has the following limitations: it is hard to achieve robustness against noisy data, which can cause a complete failure with no output in case the sequence cannot be produced by the grammar rules; it requires a high degree of domain knowledge; it is challenging to account for multiple musical dimensions in a single grammar rule; and parsing is usually so slow for long sequences that heavy pruning is necessary (CYK-parsing complexity is cubic in the length of the sequence, parallelisation does not help much, and there is no active research in developing dedicated hardware). 

\begin{figure}[t]
    \centering
    \includegraphics[width=0.9\columnwidth]{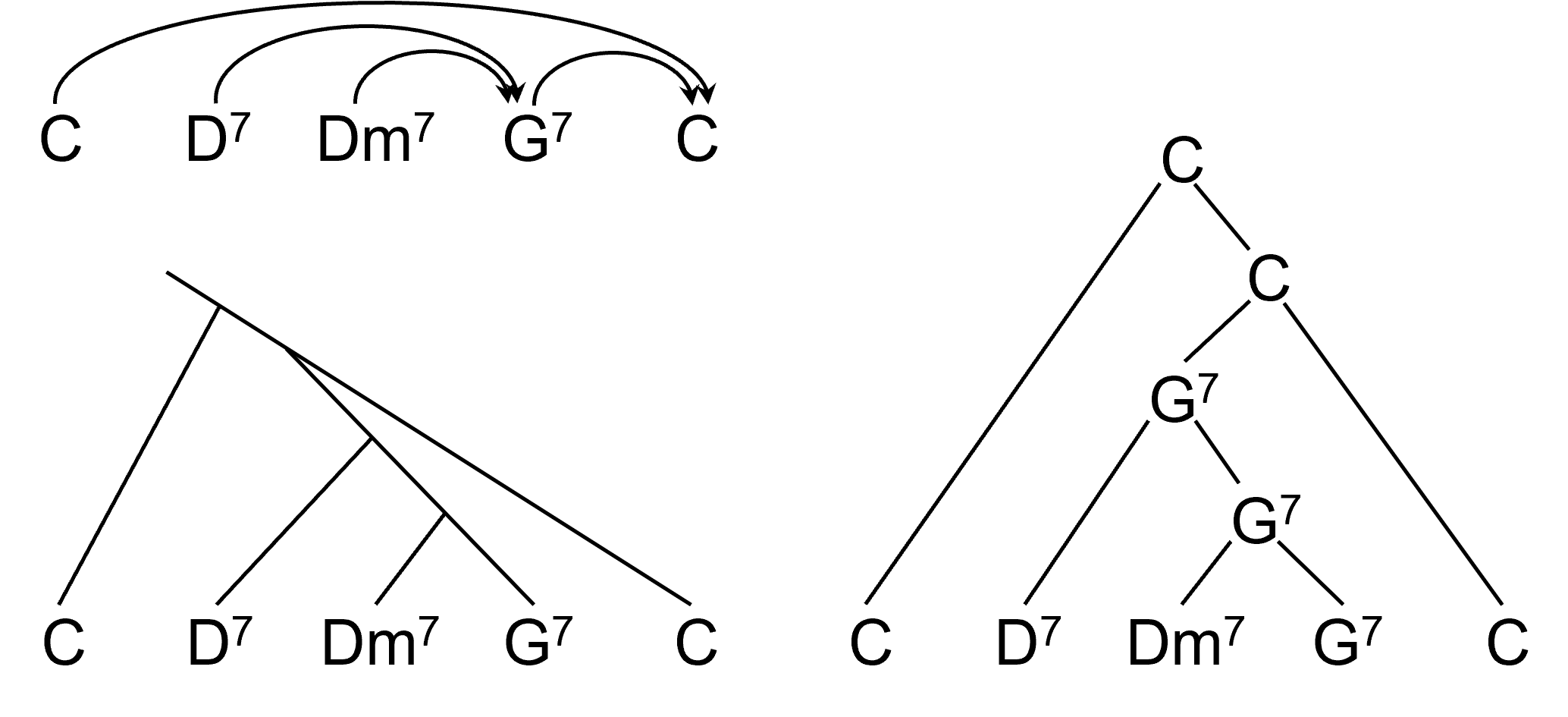}
    \caption{
    The tree harmonic analysis of the A Section of ``Take the A Train'' in three different representations. Top: dependency tree, Left: GTTM-style constituent tree. Right: CFG-style constituent tree. 
    }
    \label{fig:tree_types}
\end{figure}

Inspired by recent research in the field of natural language processing (NLP), we propose a novel, \emph{grammar-less} approach that requires little domain knowledge (only for the feature extraction phase), can easily consider multiple musical features and sequential information, produces partial results for noisy input, and is potentially scalable to longer sequences and larger datasets (since its components are proven to succeed in such scenarios). 
Our system works by predicting \textit{dependency trees} which consist of dependency arcs between the input sequence elements. Such a structure can be used as-is or later be converted into \textit{constituent trees} which are typically used to model music hierarchies (see Figure~\ref{fig:tree_types}).
The probability of each dependency arc is predicted in parallel (i.e., without considering other dependencies during prediction) by leveraging the rich contextual information produced by a transformer encoding of the input sequence. 
This set of probabilities is then run through a post-processing algorithm to ensure a valid tree structure (i.e., no cycles of dependency arcs).

We pair our Music Dependency Parser \ourapp with a procedure that enables its usage from constituent trees, and test it on two tree datasets: the time-span treebank from the GTTM database~\cite{hamanaka2014gttm}, which expresses subordinate relations between notes in monophonic melodies; and the Jazz Harmony Treebank (JHT), a set of harmonic analyses for chord sequences~\cite{harasim2020treebank}. 
We compare the results of our system with the best-performing available approaches and obtain new state-of-the-art results. 

\section{Related Work}
\textbf{Music Trees and Music Parsing. }
Trees of musical sequences have traditionally been notated as constituent trees
\cite{rohrmeier2011towards,abdallah2015analysing,nakamura2016tree,hamanakatime,granroth2014robust,harasim2018generalized, melkonian2019music,rohrmeier2020syntax,harasim2019harmonic, foscarin2019modeling,foscarin2019parse,foscarin2019diff,rohrmeier2020towards}, with few exceptions, such as the usage of a dependency-based evaluation metric~\cite{harasim2020learnability}, and the computation of pairwise voice dependencies~\cite{finkensiep2021modeling,finkensiep2023structure}. 

A system for parsing jazz chord sequences into harmonic analyses has been proposed by Harasim et al.~\cite{harasim2018generalized} and later evaluated on a larger dataset~\cite{harasim2020learnability}. We compare our results to this approach below.
Automatic grammar-based parsing of time-span GTTM trees has been attempted by Hamanaka et al.~\cite{hamanaka2006atta,hamanaka2007fatta} and Nakamura et al.~\cite{nakamura2016tree}. The latter obtained comparable results with an approach that doesn't require manual parameter tuning, and we compare our system with it. More recently, deep-learning-based approaches were also proposed~\cite{hamanaka2018deepgttm,lai2021deep,hamanakatime} but the first two focus only on GTTM metrical and grouping information, and the latter focus mainly on evaluating the usage of time-span trees for melodic morphing and we could not reproduce their results.

\noindent
\textbf{Natural Language Parsing. }
Our model architecture is inspired by the graph-based dependency parser of Dozat and Manning~\cite{dozat2016deep,Dozat2018SimplerBM}. 
This model, extended with second-order dependency predictions~\cite{wang2019second} and pretrained language models~\cite{He2019EstablishingSB}, is still the state of the art for NLP sentence parsing~\cite{zhang2020survey}. 
Still, we make some substantial changes: the embedding layer is adapted to work from musical input, the encoder is a transformer instead of an LSTM, and, instead of the bilinear layer for arc prediction, we use a linear layer. All these choices are motivated by ablation studies.

\section{Tree formats for music analyses}\label{sec:tree_formats}
In this section, we detail the types of tree used in this paper, highlight their differences, and propose algorithms to translate between them.

\subsection{Constituent vs Dependency Trees}
A tree can be defined recursively as a node with an arbitrary number (including 0) of children that are also trees. The node that is not a child of another node in the tree is called \textit{root}, the nodes that do not have children are called \textit{leaves}, and the remaining nodes are called \textit{internal nodes}.
When a tree is used to model some relations of the elements of a sequence there are two possible configurations: \textit{dependency trees}, where each node (leaf, internal, and root) represents one and only one element of the sequence; and \textit{constituent trees} where all elements of the sequence are represented in the leaves, and root and internal nodes represent nested groupings of such elements.

Among the constituent trees there exist different representations. The bottom part of Figure~\ref{fig:tree_types} shows the two kinds we consider in this paper: the one introduced by Lerdahl and Jackendoff~\cite{lerdahl1985generative} in their Generative Theory of Tonal Music (GTTM), and the one built from the Context-Free-Grammar (CFG) of jazz harmony by Harasim et al.~\cite{harasim2018generalized}. The two representations convey almost the same information: they are both binary trees (i.e., every node has either 0 or 2 children), the internal nodes are denoted by line intersections on the first, and by explicit labels on the second; they both specify an order of importance among the children (i.e., the choice of a \textit{primary} and \textit{secondary} child) by the straight line continuation, or by labelling the node with the label of the primary child. However, this latter mechanism cannot differentiate between primary and secondary when both children have the same label; therefore, the GTTM representation is slightly more informative.

\begin{figure}[t]
    \centering
    \includegraphics[width=0.9\columnwidth]{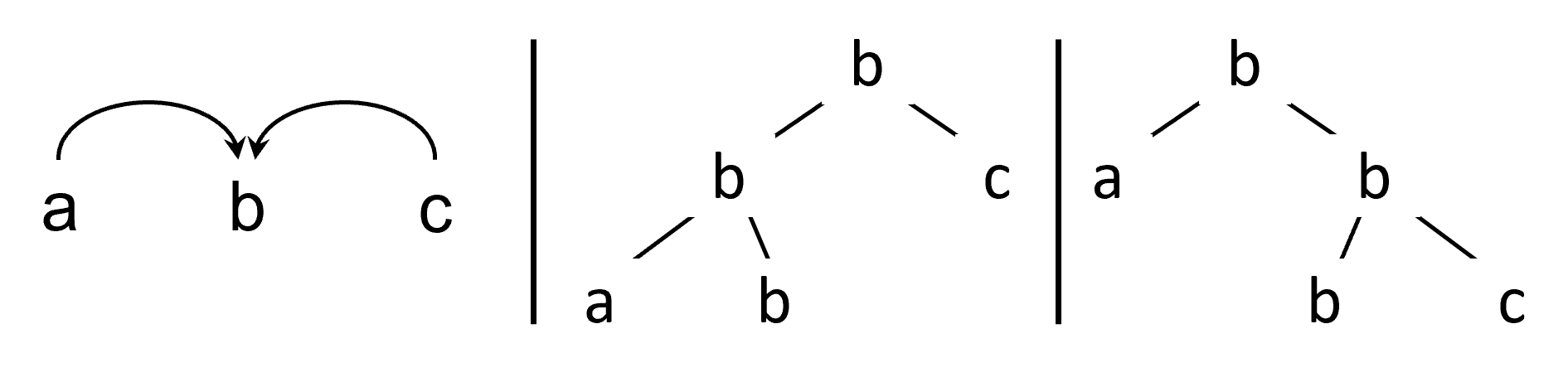}
    \caption{A dependency tree with double-sided dependencies (left). It corresponds to two possible constituent trees (middle and right).}
    \label{fig:not_unique}
\end{figure}

Our approach does not directly treat constituent trees but considers dependency trees. Each child in such a tree is called \textit{dependent}, and the node of which it is a child is called the \textit{ head}. Dependency trees can represent the same information as the binary constituent trees described above. Indeed, a dependent-head arc is equivalent to a head-labelled constituent node with two children: the primary is again the head, and the secondary is the dependent. There is only one ambiguity: the dependency tree does not encode a splitting order in the case of \textit{double-sided dependencies}, a configuration in which one head has dependents on both sides. This makes the dependency-to-constituent transformation not unique (see Figure~\ref{fig:not_unique}). This configuration is never present in our datasets (i.e., the root is always the left-most or right-most element in the sequence) thus we don't handle it. For more general datasets, one could add a binary classifier that predicts the splitting order.

The dependency trees built from the constituent trees are \textit{projective}, i.e., for all their arcs $x_\dep \rightarrow{} x_\head$, there is a path from the head to every element $x$ that lies between the head and the dependent in the sentence~\cite{stanford_book}.
This means that there are no ``crossing arcs'', e.g., $x_1 \xrightarrow{} x_3 , x_2 \xrightarrow{} x_4$.


Before proceeding with the paper, we introduce some notation we will use in the next sections.
We denote the sequence that constitutes the input of our system as $x = [x_1,\dots,x_\len]$, where $\len$ is the sequence's length. We represent the dependency tree over $x$ as the set of dependent-head\footnote{We indicate dependency arcs as arrows pointing in the direction of the head. Note that in other (NLP) papers, the opposite convention is used.} indices that corresponds to each arc $x_\dep \rightarrow{} x_\head$:
\begin{equation}
    y = \{(\dep,\head) \mid \dep,\head \in [1,\dots,\len] \}
\end{equation}

\subsection{Tree Conversion Algorithms}
Since the ground-truth annotations in our datasets are constituent trees, we translate them into dependency trees for training. We also translate tree predictions back to constituent trees to run constituent-based evaluation metrics, and when we are interested in using such a representation as input for further applications. We assume our constituent trees to be binary trees and not contain double-sided dependencies. For simplicity, we consider CFG-style constituent trees with labels in their internal nodes.

\subsubsection{Dependency to Constituent Tree}
%

%
%
Existing NLP implementations of this transformation are unnecessarily complicated for our scenario because they consider compound node labels and double-sided dependencies~\cite{kong2015transforming}.
Instead, we present a recursive top-down algorithm which yields a unique constituent solution for every single-sided dependency tree. 

The algorithm takes a fully formed dependency tree and starts with the root of the (to-be-built) constituent tree. At each step, it removes one dependency and adds two new constituent nodes. The recursive function takes as input a dependency tree node and a constituent tree node, both labelled with the same sequence element. The constituent node gets assigned two children: the primary is labelled with the element of the input nodes, and the secondary is labelled with the dependent that is further away in the sequence. The choice of which is the left and the right child respects their label position in the sequence. The considered dependency is removed from the tree and the recursive function is called two times, once for each constituent child (with the corresponding dependency node). The process stops when the dependency tree node has no dependents.

\subsubsection{Constituent to Dependency Tree}
This algorithm was used in the literature (e.g., \cite{harasim2020treebank}). It starts from a fully formed constituent tree and a dependency tree without any dependency arcs, consisting only of the nodes labelled with sequence elements. The algorithm groups all internal tree nodes with their primary child (which all have the same label) and uses all secondary child relations originating from each group to create dependency arcs between the group label and the secondary child label.

\section{Parsing Technique}
Our goal is to predict a dependency tree $y$ for a given musical sequence $x$.
Our pipeline consists of three steps: feature extraction from $x$; prediction of dependency relations; and postprocessing to ensure that the output is a valid tree structure. In the training phase, the output (before postprocessing) is compared with the ground truth dependency tree and a loss is computed to update the model parameters via backpropagation. 

\subsection{Feature Extraction}
For each input element, $x_i \in x$, we produce three groups of features. The first is a ``static'' description of the element (i.e., without any temporal information), the second encodes the element's duration, and the third encodes the element's metrical position, i.e., its position in the measure relative to the hierarchy induced by the time signature. The static description is built differently for chords and notes, while the other two are independent of the input type. Note that, due to our model architecture (see next section), we need categorical features and it is not primarily important to keep their number small or to have them ordered.

For note sequences, the \emph{static description} of each element is a single integer corresponding to either the MIDI pitch of the element in $[0,\dots,127]$ if it is a note or with the value $128$ if the element is a rest.
For chord sequences, we use three integers. The first in $[0,\dots,11]$ encodes the pitch-class of the chord root. The second in $[0,\dots,5]$ specifies the basic form of the chord among major, minor, augmented, half-diminished, diminished, and suspended (sus). The last in $[0,1,2]$ denotes a chord extension among  6, minor 7, or major 7. The chord labels were simplified by the author of the dataset to only include these extensions, but in a more general scenario, a larger set of integers could be used. The chord sequences do not contain rests.

We represent the \emph{durations of the elements} with discrete values normalised by the duration of the measure. We pre-collect the list of all durations occurring in the dataset and encode each element's duration as an index on that list. For the GTTM dataset, this would be an integer in $[0,\dots,44]$, while for the JHB dataset, it is an integer in $[0,\dots,5]$. The number of possibilities is very different, since the temporal position of chords follows much simpler rules, mostly occurring only at the beginning or in the middle of a bar for simple time signatures and at three bar positions for compound time signatures. For tied notes, we consider a single note with the total duration, and notes can last more than one measure. This is different from the annotations in the JHT in which each measure opens a new chord symbol, even if the same chord is repeated in consecutive measures.

\begin{figure}[t]
    \centering
    \includegraphics[width=\columnwidth]{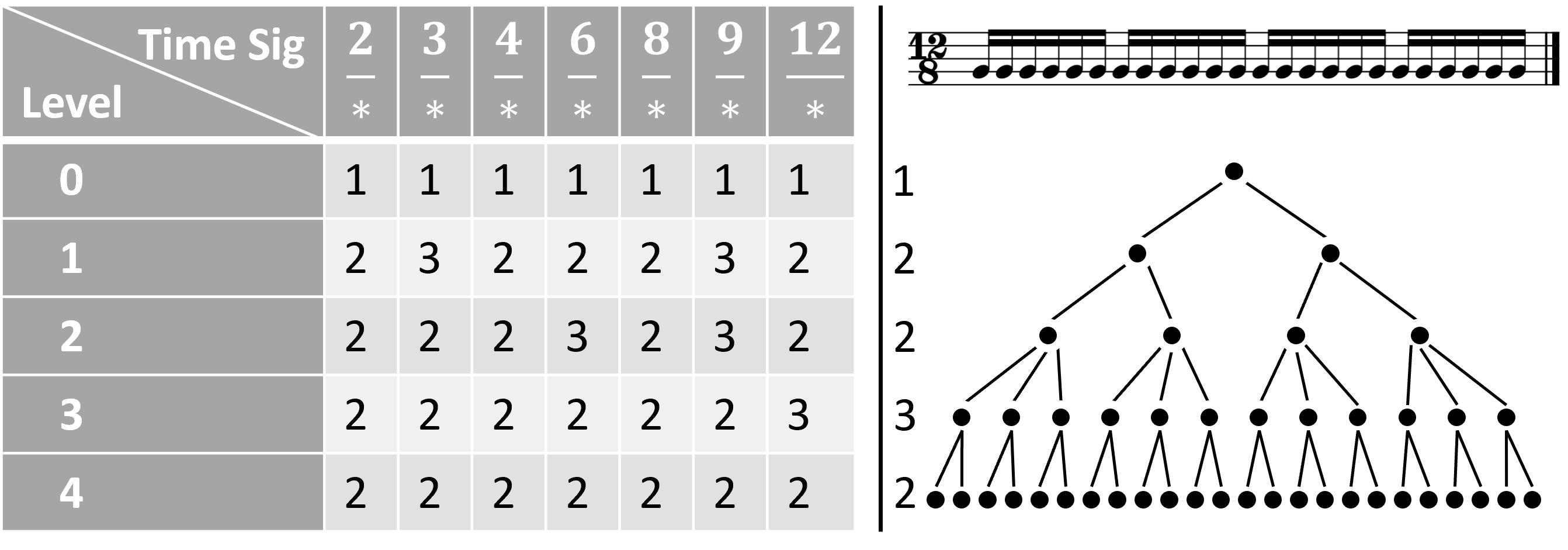}
    \caption{Left: metrical divisions $m$ for different time signature numerators. Right: visualisation of metrical divisions for a measure with time signature 12/8.}
    \label{tab:metrical division table}
\end{figure}

To represent the \emph{metrical position}, we use an inverse measure of metrical strength, encoded with a single integer in $[0,\dots, 5]$. This integer is computed as a function of the normalised temporal position in the measure $t \in [0,1[$, and the time signature numerator. Each time-signature numerator is associated with a template of metrical divisions $m$, as proposed by Foscarin~\cite{foscarin2019modeling} and here extended to more time signatures. For example, a time signature with a numerator $12$ (e.g., $12/8$ or $12/4$) will have metrical divisions $m = [1, 2, 2, 3, 2]$, i.e., the whole measure at level $0$ is divided into two parts at level $1$, each resulting part is divided  in 2 at level $2$, then 3 at level $3$, and 2 at level $4$. Table~\ref{tab:metrical division table} reports metrical divisions for all numerators we consider. 

Each level $\level$ in the metrical division defines a temporal grid with step $\delta_\level = 1/\prod_{l=0}^{\level} m_l$, and the \emph{inverse metrical strength} is defined as the lowest level for which the note position falls on the temporal grid, 
$\textrm{min}_\level(\level \mid t/ \delta_\level \in \mathbb{N})$.
For example, a time signature $6/8$ defines grids with steps 
$[1,\frac{1}{2},\frac{1}{6},\frac{1}{12},\frac{1}{24}]$, and the notes of the measure $| \musQuarter{} \musEighth{} \musQuarterDotted{}|$ will have normalised temporal position $[0,\frac{2}{6}, \frac{1}{2}]$ and inverse metrical strength $[0,2,1]$.
If the note doesn't align with any temporal grid, then its inverse metrical strength is the maximum, 5 in our settings.
Using metrical strength as input to our system may seem overly complicated. However, given the small size of our datasets and the high variety of time signatures, we need a mechanism to encode metrical information generalisable across different time signatures. It is to be expected that with a larger dataset size, this feature could be discarded, as the model could learn similar information from the list of notes with duration.

The feature extraction process lets us build the input matrix $\X \in \mathbb{N}^{\len \times \nfeat}$ where $\len$ is the sequence length, and $\nfeat$ is the number of features for each element: 3 for the GTTM dataset, and 5 for the JHT dataset. Before moving on, it has to be noted that there exist other more general ways of transforming symbolic music into convenient inputs for deep learning models, notably the tokenisation techniques, e.g.,\cite{miditok2021,fradet2023byte} inspired by NLP research. However, given the small dimension of our dataset and the fact that our melodies are strictly monophonic, we prefer to use a more compact, ad-hoc input representation. Our parsing framework remains general and usable with other techniques.

\subsection{Model}
\begin{figure}[t]
    \centering
    \includegraphics[width=\columnwidth]{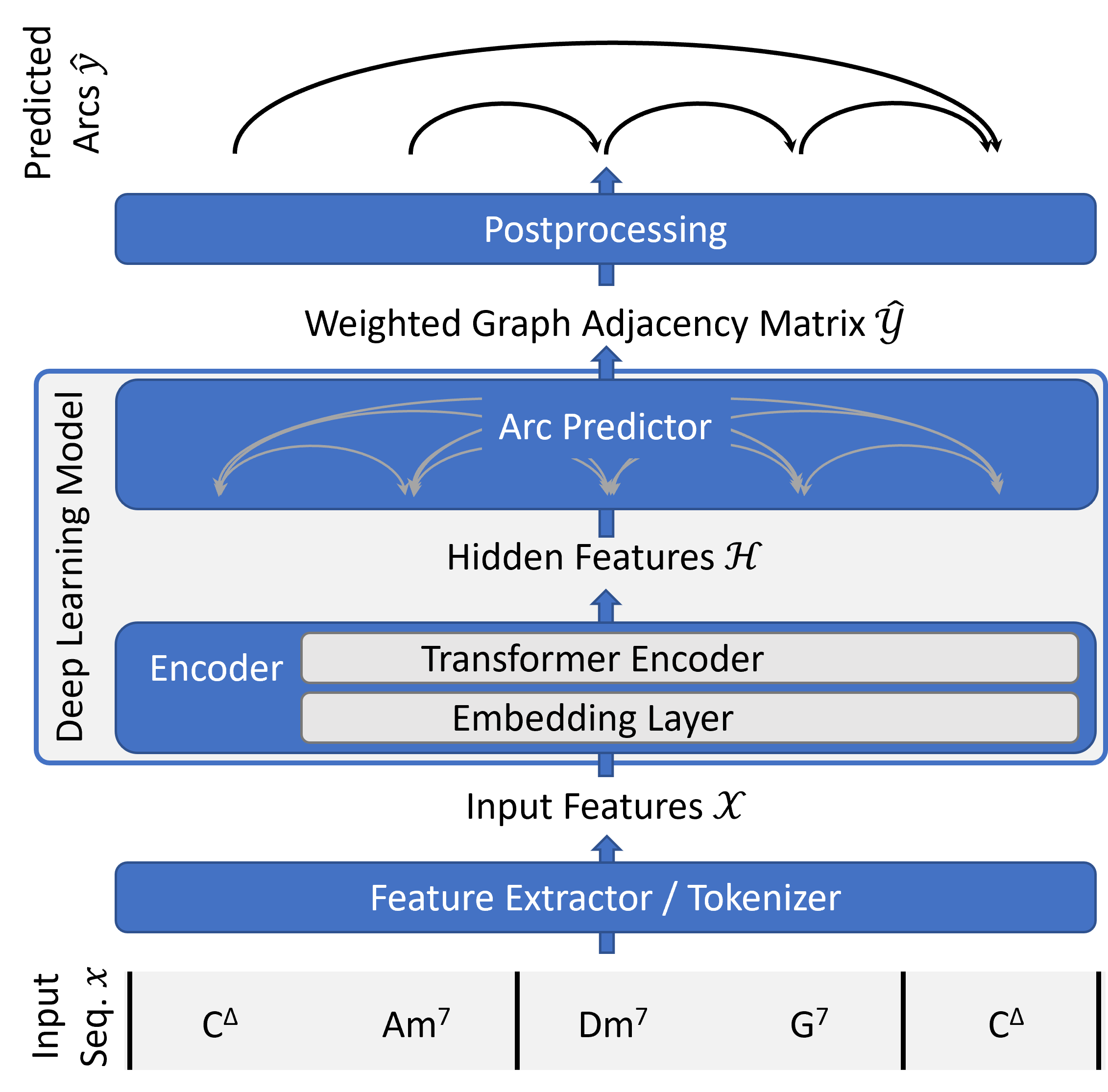}
    \caption{Architecture of our model. The input displayed is an example of a chord sequence, but the same architecture is used for note sequences.}
    \label{fig:architecture}
\end{figure}

Our model consists of two parts: an encoder and an arc predictor (see Figure~\ref{fig:architecture}). The goal of the encoder is to enrich the input features $\X$ with contextual information. The arc predictor uses the enriched sequence features to predict whether each possible pair of elements in the input sequence should be connected by a dependency arc.

The first part of the encoder is an embedding layer, a learnable look-up table which maps our collection of categorical features (each integer) to points in a continuous multidimensional space. Specifically, we use $\nfeat$ embedding layers (one for each input feature), which work independently, and map all values that the feature can have to a vector of a fixed embedding dimension. All vectors are then summed together to obtain a unique representation while keeping the input size small (see \cite{sum_emb} for an explanation of why summing is better than concatenating). 
After the embedding layer, we have the encoder part of a transformer~\cite{vaswani2017attention} with relative position representations~\cite{Shaw2018SelfAttentionWR}. It outputs a matrix with the same number of rows as the input matrix $\X$ (one for each sequence element) but with a (possibly) different number of hidden-feature columns $h$. Onto this, we concatenate a new learnable single row that acts as the head of the root node.  The result is a new matrix $\Hid \in \mathbb{Q}^{(\len +1) \times h}$.

The arc predictor part of our model is a multilayer perceptron (MLP) that performs the binary classification task of deciding whether each pair $(x_\head,x_\dep)$ in the Cartesian product of the input elements, i.e., $\{(\head,\dep) \mid \forall \ \head, \dep \in [1,\dots,\len]\}$, should be connected by a dependency arc.
Depending on the input representation and the specific task we are targeting, there may be some pairs that are not connectable by a dependency arc, for example, pairs where $\head = \dep$. For the GTTM input, pairs for which at least one element is a rest are also not connectable. Therefore, the binary classification is performed only on a subset of all pairs $\potarcs$ that we call \textit{potential arcs}.
For every potential arc $(x_\dep,x_\head)$, we predict the probability $\hat{y}_{\dep, \head}$ of a dependency arc by concatenating the two rows of $\Hid$ that correspond to the head's and the dependent's index into a single vector of length $2h$ and giving it as input to the MLP.
We concatenate the two inputs instead of summing or multiplying them because our arcs are directed, so we need to preserve the order when aggregating the two embeddings. Moreover, despite the bilinear layer being a major selling point of Dozat's paper~\cite{dozat2016deep}, we find that the concatenation approach yields better results.
We can collect the output for all potential arcs into a \textit{weighted graph-adjacency matrix} $\hat{Y}$, which is a $\len \times \len$ matrix with entries $\hat{y}_{\head, \dep}$ at the corresponding indices. We assign a probability 0 to the matrix entries that correspond to arcs $\notin \potarcs$.

\subsection{Training Loss}\label{subsec:loss}

In the training phase, we use the sum of the binary-cross-entropy (BCE) loss and the (multiclass) cross-entropy (CE) loss. The BCE loss is computed independently for each potential arc and measures the difference between the ground-truth label (0 or 1) and the predicted probability. We also use the CE loss because our problem can be framed as a multiclass classification problem where for each element we predict his head among $\len + 1$ possibilities (each sequence element plus a dummy element for the root and rests elements). The CE loss is therefore applied column-wise to the adjacency matrix $\hat{Y}$ predicted by our model.

In NLP, the BCE loss was used by \cite{Dozat2018SimplerBM} while the CE loss is used more generally, for example, by~\cite{dozat2016deep,fernandez2020transition}. We experimentally found that the sum of the two losses yields the best results.

\subsection{Postprocessing}
Since the prediction of our model is made independently for each potential arc, simply taking the row-wise maximum of the weighted adjacency matrix to select which head to assign to each element of the sequence could produce dependency cycles and, therefore, not yield a tree structure.
We use a maximum-spanning-tree algorithm to find the tree over $\hat{Y}$ with the highest weight. Since our dependency trees are projective, we use the Eisner algorithm~\cite{eisner-1996-three} which solves this problem using bottom–up dynamic programming with a time complexity of $O(\len^3)$.
For applications involving non-projective trees other post-processing approaches such as Chu-Liu/Edmonds~\cite{edmonds1967optimum,chu1965shortest} ($O(\len^2)$) are implemented in our framework.

\section{Experiments}
Below, we describe the datasets, evaluation metrics, and experimental settings for the two kinds of trees we consider.

\subsection{Datasets and preprocessing}
We obtain the melodic time-span trees from the GTTM database~\cite{hamanaka2014gttm}, which contains MusicXML encodings of monophonic scores and a dedicated XML-based encoding of the constituent time-span trees (among other trees that we don't consider in this paper). We extract the note features with the Python library Partitura~\cite{partitura_mec}. Some pieces have two different trees, and we keep only the first. We also discard 4 pieces that we could not import due to inconsistencies in the XML file encoding. In total, we have 296 melodies of lengths between 10 and 127 (notes + rests). For training, we augment the dataset by considering all transpositions between one octave higher and one octave lower.

We obtain the chord analyses from the Jazz Harmony Treebank (JHT)~\cite{harasim2019harmonic}, which encodes both chord labels and harmonic analyses as constituent trees, in JSON format. As discussed in Section~\ref{sec:tree_formats} this format does not distinguish between the primary and the secondary child when both have the same chord label. In this case, we assume by default that the right is the primary. The dataset contains two kinds of trees: open and complete constituents. We use the former for comparison reasons since the results are reported only for those~\cite{harasim2020learnability}. In total, we have 150 sequences of lengths between 11 and 38 chords. For training, we augment the dataset by considering all 12 possible transpositions of the chord roots.

\subsection{Evaluation metrics}
The papers we compare use different metrics, and we implement all of them. The work of Harasim~\cite{harasim2020learnability} uses two metrics, one more relevant for dependency trees and the other for constituent trees. The first is the \textit{arc accuracy}, i.e., the normalised cardinality of the intersection between the set of predicted arcs and the ground truth arcs. The second is the \textit{span accuracy}, computed as the normalised cardinality of the intersection between all the spans of the predicted constituent tree (i.e., the pair of the leftmost and rightmost leaf that is part of the subtree rooted at any non-leaf node) and the spans of the ground truth tree (see \cite{harasim2020learnability} for a more detailed explanation). Nakamura et al.~\cite{nakamura2016tree} use the \textit{node accuracy} metric, i.e., the normalised cardinality of the intersection between nodes in the predicted and ground truth trees, where two nodes are considered equal if the labels of the parent and children (or a dummy label if the node have no parent or children) are equal.

We also report another metric, the \textit{head accuracy}, computed as the multiclass classification accuracy on the indices of the predicted heads, ordered by their dependent. For example for the dependency tree of Figure~\ref{fig:tree_types}, this would correspond to the accuracy computed on the sequence $[4,2,3,4,-1]$, where $-1$ indicates the root (which has no head). This is similar to the arc accuracy but enforces the presence of a dependency head for each sequence element (which may not be the case for a generic system), and gives more weight to the correct root prediction. It is also faster to compute and commonly used in NLP, so we include it to set a metric for future research.
Note that all metrics presented above don't consider the nodes corresponding to rests, since they are only part of the input sequence, but not part of the tree.

\subsection{Results}
For our experiments, we set the hyperparameters of our encoder to an embedding size of 96, and 2 transformer layers of hidden size 64. The arc predictor (MLP) has 2 linear layers with the same hidden size.
We use the GeLU activation~\cite{gelu} and the AdamW optimiser~\cite{adamw}, with a learning rate of 0.0004 and weight decay of 0.05. We train with learning rate warm-up~\cite{huang2020improving} of 50 steps and cosine annealing to limit the problem of high variance in the initial and final stages of training. The latter was particularly important since we did not use a validation set to perform early stopping due to the small size of our datasets. 
We train for 60 and 20 epochs for the JHT and GTTM datasets, respectively, since the latter is bigger and the data augmentation yields twice as many pieces in total). The training time is roughly the same, around 1 hour on a GPU RTX 1080.

\begin{figure}[t]
    \centering
    \includegraphics[width=\columnwidth]{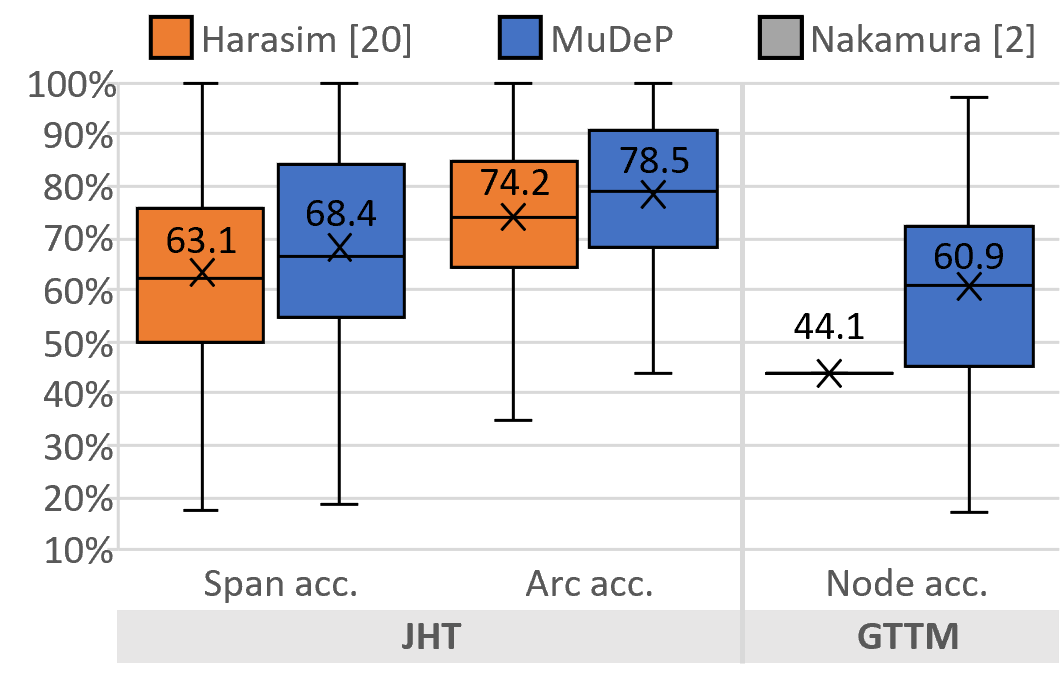}
    \caption{
    Boxplots of three accuracy metrics (higher is better) computed with leave-one-out cross-validation and their average. For Nakamura et al.~\cite{nakamura2016tree}, we report the average from their paper, so there is no deviation information.
    }
    \label{fig:main_results}
\end{figure}

We compare the results of our \ourapp on the JHT with Harasim~\cite{harasim2020learnability} and on the GTTM with Nakamura et al.~\cite{nakamura2016tree}.
We use leave-one-out cross-validation, i.e., for a dataset with $N$ pieces, we run our system $N$ times, by training on $N-1$ pieces and evaluating with the remaining one. 
As shown in Figure~\ref{fig:main_results}, \ourapp outperforms previous methods.

By comparing the head accuracy between JHT and GTTM (79.2\% vs 57.9\%), it is clear that time-span prediction is a much harder problem than the chord analysis problem, despite the dataset being bigger. 
Another interesting result is that the span accuracy is lower than head accuracy for the JHT dataset (63.1\%), but higher for the GTTM(64.8\%). Apparently, the main problem for JHT is to select which two chords to connect, but the arc direction (i.e., which is the head and which is the dependent) is almost always correctly inferred; conversely, for the GTTM dataset, the system often connects the correct notes, but in the wrong direction. And this type of misprediction is punished in the head accuracy, but not in the span accuracy.

The full piece-wise statistics on all metrics, a graphical rendering of all our predicted trees, and the qualitative evaluation of some examples are available in our repository.


\subsection{Ablation study}
We report the difference in head accuracy averaged over 10 runs with 90/10 random train/test split for the JHT dataset.
Regarding the loss, sole usage of the (multiclass) CE loss reduced the accuracy by 0.3\%, and only using the binary CE loss reduced the accuracy by 4.1\%. 
The use of a bilinear layer in the decoder reduced the accuracy by 1.2\%. 
The absence of post-processing did not reduce the accuracy (when the network is fully trained, otherwise the reduction is very evident). This is promising but it does not automatically implies that the network is producing correctly formed trees since dependency loops could still be present in the output. There are also cases when the postprocessing is reducing the accuracy, by incorrectly deciding which arc to remove in a dependency loop.

\section{Conclusion and Future Work}
\label{sec:Conclusion and Future Work}
We presented \ourapp, a system for the dependency parsing of music sequences, and a procedure to make it applicable to constituent trees. \ourapp improves upon previous methods, by incorporating the ability to consider multiple musical features simultaneously, taking advantage of sequential context, and handling noisy inputs robustly. Moreover, since it is based on widely researched deep learning components, it has the potential to scale to large datasets and longer sequences. The bottleneck for such scalability is the postprocessing algorithm with cubic complexity. Two solutions exist to this problem: if one is interested in non-projective trees, algorithms with a square complexity are available. Apart from that, our system is already having good accuracy without the postprocessing phase, as highlighted in the ablation study. Therefore, a faster heuristic may suffice to correct the few problematic dependencies without decreasing the performance.

Since our deep learning model is a black box, it is notably complicated to find 
a human-understandable explanation of its functioning. Although work in this direction exists~\cite{mishra2017local,foscarin2022concept}, it is still very limited~\cite{praher2021veracity}. Therefore, our model is mainly intended for scenarios in which one is only interested in obtaining the parsing trees, for example, to use them as input for another MIR task.
Conversely, this paper might have limited utility if one's goal is to model music understanding and interpretation by humans. Grammar-based models are much more suitable for this goal, although there is a (somewhat speculative) possibility that the dependency-arc probabilities in our approach relate to first-guess heuristics.

As research on the deep learning components we use is rapidly evolving, any new discovery is likely to benefit our system. Self-supervised pretraining on larger datasets of monophonic music or chord sequences, for example by predicting next or masked tokens, could also improve the performance, as already proved for language parsing. While the goal of this paper was to present a general framework, we can also think about several domain-specific improvements, for example, training the GTTM time-span parser with a multi-target approach to predict at the same time the metrical, time-span, and prolongation structure. We hope that this work will motivate the development of more datasets of hierarchical music analyses, including datasets of dependency trees, which may be a valid alternative to constituent structures, and even open up more possibilities due to the missing projectivity constraints. Finally, we intend to explore in future research how the knowledge encoded in our model could be reused to guide other tasks, for example, automatic chord recognition from audio files.


\section{Acknowledgements}

This work is supported by the European Research Council (ERC) under the EU’s Horizon 2020 research \& innovation programme, grant agreement No.~101019375 (“Whither Music?”), and the Federal State of Upper Austria (LIT AI Lab).

\bibliography{biblio}

\begin{thebibliography}{10}
\providecommand{\url}[1]{#1}
\csname url@samestyle\endcsname
\providecommand{\newblock}{\relax}
\providecommand{\bibinfo}[2]{#2}
\providecommand{\BIBentrySTDinterwordspacing}{\spaceskip=0pt\relax}
\providecommand{\BIBentryALTinterwordstretchfactor}{4}
\providecommand{\BIBentryALTinterwordspacing}{\spaceskip=\fontdimen2\font plus
\BIBentryALTinterwordstretchfactor\fontdimen3\font minus
  \fontdimen4\font\relax}
\providecommand{\BIBforeignlanguage}[2]{{%
\expandafter\ifx\csname l@#1\endcsname\relax
\typeout{** WARNING: IEEEtran.bst: No hyphenation pattern has been}%
\typeout{** loaded for the language `#1'. Using the pattern for}%
\typeout{** the default language instead.}%
\else
\language=\csname l@#1\endcsname
\fi
#2}}
\providecommand{\BIBdecl}{\relax}
\BIBdecl

\bibitem{abdallah2015analysing}
S.~Abdallah, N.~Gold, and A.~Marsden, ``Analysing symbolic music with
  probabilistic grammars,'' \emph{Computational music analysis}, pp. 157--189,
  2015.

\bibitem{nakamura2016tree}
E.~Nakamura, M.~Hamanaka, K.~Hirata, and K.~Yoshii, ``Tree-structured
  probabilistic model of monophonic written music based on the generative
  theory of tonal music,'' in \emph{Proceedings of the International Conference
  on Acoustics, Speech and Signal Processing (ICASSP)}.\hskip 1em plus 0.5em
  minus 0.4em\relax IEEE, 2016, pp. 276--280.

\bibitem{hamanakatime}
M.~Hamanaka, K.~Hirata, and S.~Tojo, ``Time-span tree leveled by duration of
  time-span,'' in \emph{Proceedings of the International Symposium on Computer
  Music Multidisciplinary Research (CMMR)}, 2021, pp. 155--164.

\bibitem{finkensiep2021modeling}
C.~Finkensiep and M.~A. Rohrmeier, ``Modeling and inferring proto-voice
  structure in free polyphony,'' in \emph{Proceedings of the International
  Society for Music Information Retrieval Conference ({ISMIR})}, 2021, pp.
  189--196.

\bibitem{rohrmeier2011towards}
M.~Rohrmeier, ``Towards a generative syntax of tonal harmony,'' \emph{Journal
  of Mathematics and Music}, vol.~5, no.~1, pp. 35--53, 2011.

\bibitem{granroth2014robust}
M.~Granroth-Wilding and M.~Steedman, ``A robust parser-interpreter for jazz
  chord sequences,'' \emph{Journal of New Music Research}, vol.~43, no.~4, pp.
  355--374, 2014.

\bibitem{harasim2018generalized}
D.~Harasim, M.~Rohrmeier, and T.~J. O'Donnell, ``A generalized parsing
  framework for generative models of harmonic syntax.'' in \emph{Proceedings of
  the International Society for Music Information Retrieval Conference
  ({ISMIR})}, 2018, pp. 152--159.

\bibitem{melkonian2019music}
O.~Melkonian, ``Music as language: putting probabilistic temporal graph
  grammars to good use,'' in \emph{Proceedings of the ACM SIGPLAN International
  Workshop on Functional Art, Music, Modeling, and Design}, 2019, pp. 1--10.

\bibitem{harasim2019harmonic}
D.~Harasim, T.~J. O’Donnell, and M.~A. Rohrmeier, ``Harmonic syntax in time:
  rhythm improves grammatical models of harmony,'' in \emph{Proceedings of the
  International Society for Music Information Retrieval Conference ({ISMIR})},
  2019, pp. 335--342.

\bibitem{foscarin2019modeling}
F.~Foscarin, F.~Jacquemard, and P.~Rigaux, ``Modeling and learning rhythm
  structure,'' in \emph{Proceedings of the Sound and Music Computing Conference
  (SMC)}, 2019.

\bibitem{foscarin2019parse}
F.~Foscarin, F.~Jacquemard, P.~Rigaux, and M.~Sakai, ``A parse-based framework
  for coupled rhythm quantization and score structuring,'' in \emph{Proceedings
  of the Mathematics and Computation in Music International Conference
  (MCM)}.\hskip 1em plus 0.5em minus 0.4em\relax Springer, 2019, pp. 248--260.

\bibitem{foscarin2019diff}
F.~Foscarin, R.~Fournier-S’Niehotta, and F.~Jacquemard, ``A diff procedure
  for xml music score files,'' in \emph{Proceedings of the International
  Conference on Digital Libraries for Musicology (DLfM)}, 2019.

\bibitem{rohrmeier2020towards}
M.~Rohrmeier, ``Towards a formalization of musical rhythm.'' in
  \emph{Proceedings of the International Society for Music Information
  Retrieval Conference ({ISMIR})}, 2020, pp. 621--629.

\bibitem{fitch2014hierarchical}
W.~T. Fitch and M.~D. Martins, ``Hierarchical processing in music, language,
  and action: Lashley revisited,'' \emph{Annals of the New York Academy of
  Sciences}, vol. 1316, no.~1, pp. 87--104, 2014.

\bibitem{tsuchiya2013probabilistic}
M.~Tsuchiya, K.~Ochiai, H.~Kameoka, and S.~Sagayama, ``Probabilistic model of
  two-dimensional rhythm tree structure representation for automatic
  transcription of polyphonic midi signals,'' in \emph{Proceedings of the
  Asia-Pacific Signal and Information Processing Association Annual Summit and
  Conference}.\hskip 1em plus 0.5em minus 0.4em\relax IEEE, 2013, pp. 1--6.

\bibitem{sakai1961syntax}
I.~Sakai, ``Syntax in universal translation,'' in \emph{Proceedings of the
  International Conference on Machine Translation and Applied Language
  Analysis}, 1961.

\bibitem{hamanaka2014gttm}
M.~Hamanaka, K.~Hirata, and S.~Tojo, ``Musical structural analysis database
  based on gttm,'' in \emph{Proceedings of the International Society for Music
  Information Retrieval Conference ({ISMIR})}, 2014, pp. 325--330.

\bibitem{harasim2020treebank}
D.~Harasim, C.~Finkensiep, P.~Ericson, T.~J. O'Donnell, and M.~Rohrmeier, ``The
  jazz harmony treebank,'' in \emph{Proceedings of the International Society
  for Music Information Retrieval Conference ({ISMIR})}, 2020, pp. 207--215.

\bibitem{rohrmeier2020syntax}
M.~Rohrmeier, ``The syntax of jazz harmony: diatonic tonality, phrase
  structure, and form,'' \emph{Music Theory and Analysis (MTA)}, vol.~7, no.~1,
  pp. 1--63, 2020.

\bibitem{harasim2020learnability}
D.~Harasim, ``The learnability of the grammar of jazz: Bayesian inference of
  hierarchical structures in harmony,'' Ph.D. dissertation, EPFL, 2020.

\bibitem{finkensiep2023structure}
C.~Finkensiep, ``The structure of free polyphony,'' Ph.D. dissertation, EPFL,
  2023.

\bibitem{hamanaka2006atta}
M.~Hamanaka, K.~Hirata, and S.~Tojo, ``Implementing “a generative theory of
  tonal music”,'' \emph{Journal of New Music Research}, vol.~35, no.~4, pp.
  249--277, 2006.

\bibitem{hamanaka2007fatta}
------, ``{FATTA}: Full automatic time-span tree analyzer,'' in
  \emph{International Computer Music Conference (ICMC)}, vol.~1, 2007, pp.
  153--156.

\bibitem{hamanaka2018deepgttm}
------, ``{deepGTTM-III: Multi-task Learning with Grouping and Metrical
  Structures},'' in \emph{Proceedings of the International Symposium on
  Computer Music Multidisciplinary Research (CMMR)}, 2018, pp. 238--251.

\bibitem{lai2021deep}
Y.-R. Lai and A.~W.-Y. Su, ``Deep learning based detection of {GPR6 GTTM}
  global feature rule of music scores,'' in \emph{Proceedings of the
  International Conference on New Music Concepts}, vol.~56, 2021.

\bibitem{dozat2016deep}
T.~Dozat and C.~D. Manning, ``Deep biaffine attention for neural dependency
  parsing,'' in \emph{Proceedings of the International Conference on Learning
  Representations (ICLR)}, 2017.

\bibitem{Dozat2018SimplerBM}
------, ``Simpler but more accurate semantic dependency parsing,'' in
  \emph{Proceedings of the Annual Meeting of the Association for Computational
  Linguistics}, 2018.

\bibitem{wang2019second}
X.~Wang, J.~Huang, and K.~Tu, ``Second-order semantic dependency parsing with
  end-to-end neural networks,'' 2019, pp. 4609–--4618.

\bibitem{He2019EstablishingSB}
H.~He and J.~D. Choi, ``Establishing strong baselines for the new decade:
  Sequence tagging, syntactic and semantic parsing with bert,'' in
  \emph{Proceedings of the International Florida Artificial Intelligence
  Research Society Conference}, 2019, pp. 228--233.

\bibitem{zhang2020survey}
M.~Zhang, ``A survey of syntactic-semantic parsing based on constituent and
  dependency structures,'' \emph{Science China Technological Sciences},
  vol.~63, no.~10, pp. 1898--1920, 2020.

\bibitem{lerdahl1985generative}
F.~Lerdahl and R.~S. Jackendoff, \emph{A generative theory of tonal
  music}.\hskip 1em plus 0.5em minus 0.4em\relax MIT press, 1985.

\bibitem{stanford_book}
J.~Daniel, M.~James~H \emph{et~al.}, \emph{Speech and language processing: An
  introduction to natural language processing, computational linguistics, and
  speech recognition}.\hskip 1em plus 0.5em minus 0.4em\relax Prentice Hall
  series in artificial intelligence, 2007.

\bibitem{kong2015transforming}
L.~Kong, A.~M. Rush, and N.~A. Smith, ``Transforming dependencies into phrase
  structures,'' in \emph{Proceedings of the Conference of the North American
  Chapter of the Association for Computational Linguistics: Human Language
  Technologies}, 2015, pp. 788--798.

\bibitem{miditok2021}
N.~Fradet, J.-P. Briot, F.~Chhel, A.~El~Fallah~Seghrouchni, and N.~Gutowski,
  ``{MidiTok}: A python package for {MIDI} file tokenization,'' in
  \emph{Late-Breaking Demo Session of the International Society for Music
  Information Retrieval Conference (ISMIR)}, 2021.

\bibitem{fradet2023byte}
N.~Fradet, J.-P. Briot, F.~Chhel, A.~E.~F. Seghrouchni, and N.~Gutowski, ``Byte
  pair encoding for symbolic music,'' \emph{arXiv preprint arXiv:2301.11975},
  2023.

\bibitem{sum_emb}
{EuroCC National Competence Center Sweden (ENCCS)}, ``Graph neural networks and
  transformer workshop,''
  \url{https://enccs.github.io/gnn_transformers/notebooks/session_1/1b_vector_sums_vs_concatenation/},
  2022.

\bibitem{vaswani2017attention}
A.~Vaswani, N.~Shazeer, N.~Parmar, J.~Uszkoreit, L.~Jones, A.~N. Gomez,
  {\L}.~Kaiser, and I.~Polosukhin, ``Attention is all you need,''
  \emph{Advances in neural information processing systems}, vol.~30, 2017.

\bibitem{Shaw2018SelfAttentionWR}
P.~Shaw, J.~Uszkoreit, and A.~Vaswani, ``Self-attention with relative position
  representations,'' in \emph{Proceedings of the North American Chapter of the
  Association for Computational Linguistics}, 2018.

\bibitem{fernandez2020transition}
D.~Fern{\'a}ndez-Gonz{\'a}lez and C.~G{\'o}mez-Rodr{\'\i}guez,
  ``Transition-based semantic dependency parsing with pointer networks,''
  \emph{Proceedings of the Annual Meeting of the Association for Computational
  Linguistics}, 2020.

\bibitem{eisner-1996-three}
J.~M. Eisner, ``Three new probabilistic models for dependency parsing: An
  exploration,'' in \emph{Proceedings of the International Conference on
  Computational Linguistics ({COLING})}, 1996.

\bibitem{edmonds1967optimum}
J.~Edmonds, ``Optimum branchings,'' \emph{Journal of Research of the national
  Bureau of Standards}, vol.~71, no.~4, pp. 233--240, 1967.

\bibitem{chu1965shortest}
Y.-J. Chu, ``On the shortest arborescence of a directed graph,'' \emph{Scientia
  Sinica}, vol.~14, pp. 1396--1400, 1965.

\bibitem{partitura_mec}
C.~E. Cancino-Chac\'{o}n, S.~D. Peter, E.~Karystinaios, F.~Foscarin,
  M.~Grachten, and G.~Widmer, ``{Partitura: A Python Package for Symbolic Music
  Processing},'' in \emph{{Proceedings of the Music Encoding Conference
  (MEC)}}, Halifax, Canada, 2022.

\bibitem{gelu}
D.~Hendrycks and K.~Gimpel, ``Gaussian error linear units {(GELUs)},''
  \emph{arXiv preprint arXiv:1606.08415}, 2016.

\bibitem{adamw}
I.~Loshchilov and F.~Hutter, ``Decoupled weight decay regularization,'' in
  \emph{International Conference on Learning Representations (ICLR)}, 2019.

\bibitem{huang2020improving}
X.~S. Huang, F.~Perez, J.~Ba, and M.~Volkovs, ``Improving transformer
  optimization through better initialization,'' in \emph{Proceedings of the
  International Conference on Machine Learning (ICML)}, 2020, pp. 4475--4483.

\bibitem{mishra2017local}
S.~Mishra, B.~L. Sturm, and S.~Dixon, ``{Local Interpretable Model-agnostic
  Explanations for Music Content Analysis},'' in \emph{Proceedings of the
  International Society for Music Information Retrieval Conference ({ISMIR})},
  2017, pp. 537--543.

\bibitem{foscarin2022concept}
F.~Foscarin, K.~Hoedt, V.~Praher, A.~Flexer, and G.~Widmer, ``Concept-based
  techniques for "musicologist-friendly" explanations in a deep music
  classifier,'' in \emph{Proceedings of the International Society for Music
  Information Retrieval Conference ({ISMIR})}, 2022.

\bibitem{praher2021veracity}
V.~Praher, K.~Prinz, A.~Flexer, and G.~Widmer, ``On the veracity of local,
  model-agnostic explanations in audio classification: targeted investigations
  with adversarial examples,'' in \emph{Proceedings of the International
  Society for Music Information Retrieval Conference ({ISMIR})}, 2021.

\end{thebibliography}

%
%
%
%
%

\end{document}